\documentclass[twocolumn]{IEEEtran} 

\usepackage[dvips]{graphicx}

\def\BibTeX{{\rm B\kern-.05em{\sc i\kern-.025em b}\kern-.08em
    T\kern-.1667em\lower.7ex\hbox{E}\kern-.125emX}}

\begin{document}

\title{A Self-Reconfigurable Computing Platform Hardware Architecture} 

\author{Andreas Weisensee, Darran Nathan\thanks{The authors are with the
DSP Technology Centre, School of Engineering, NgeeAnn Polytechnic, Singapore. (e-mail: andreas@projectproteus.org [Andreas Weisensee]).}}

\markboth{Project Proteus}
{Murray and Balemi: Using the Document Class IEEEtran.cls} 

\maketitle

\begin{abstract}
Field Programmable Gate Arrays (FPGAs) have recently been increasingly used for highly-parallel processing of compute intensive tasks. This paper introduces an FPGA hardware platform architecture that is PC-based, allows for fast reconfiguration over the PCI bus, and retains a simple physical hardware design. The design considerations are first discussed, then the resulting system architecture designed is illustrated. Finally, experimental results on the FPGA resources utilized for this design are presented.
\end{abstract}


\section{Introduction}
\label{sectIntroduction}

Computer processors have for many years been designed based on the von-Neumann or Harvard architectures. Software to be run on these processors are compiled into a set of processor-specific instructions, which are loaded during run-time and executed sequentially. Such sequential processing of an instruction every few clock cycles works well enough for typical PC applications such as text editors, which have low data processing requirement.

However, PCs are also often used for computationally intensive high-throughput data processing, especially in scientific research work. The sequential nature of the typical PC processor, such as the Intel Pentium, becomes a major processing bottleneck in such situations. The solution to this problem has been to use processors with greater clockspeeds, or to network several of these PCs together into a cluster or computational grid \cite{bibGlobus}.

More recently, there has been an increasing interest in the use of reconfigurable hardware chips for such compute intensive data processing. These chips, such as Field Programmable Gate Arrays (FPGAs), possess a fundamentally different architecture from the typical von-Neumann or Harvard type processors. The algorithms to be executed are normally defined in a hardware description language and compiled into a bitstream, which will be downloaded to the FPGA as and when use of the algorithm is desired. This bitstream download will reconfigure the hardware logic on the FPGA accordingly, allowing data passed into the FPGA to be processed in hardware, in parallel.

Several reconfigurable computing research projects \cite{bibRawMIT} \cite{bibPipeRench} \cite{bibGarp} focus on developing new, improved designs of reconfigurable chips. Other groups \cite{bibUMASS} \cite{bibToronto} \cite{bibBYU} utilize off-the-shelf FPGAs, such as those from Xilinx \cite{bibXilinx}, and work on issues such as logic placement and routing optimization \cite{bibVPR}. 

Project Proteus \cite{bibProjectProteus} was initiated by the DSP Technology Centre of NgeeAnn Polytechnic to develop a low-cost FPGA-based reconfigurable computing platform for typical PCs, with a portable software platform layer and using off-the-shelf hardware components. The hardware is designed for fast FPGA reconfiguration operations, with minimal physical hardware component count and complexity, while maintaining the desirable features of a reconfigurable platform such as configuration bitstream readback and dynamic reconfigurability features. This paper discusses the requirements and design of this hardware architecture.

Section \ref{sectDesignConsiderations} describes the requirements of the reconfigurable computing platform hardware architecture and how this compares to existing solutions, Section \ref{sectSystemArchitecture} discusses the architectural issues addressed and the design of the hardware, Section \ref{sectExperiments} presents some experimental results on the FPGA resource requirements, and finally Section \ref{sectConclusion} concludes this paper.

\section{Design Considerations}
\label{sectDesignConsiderations}
To understand the architecture of the hardware, it will be useful to first discuss the requirements imposed by its intended use and desired features.

Firstly, the reconfigurable computing platform is intended to be PC-based. The FPGA will therefore be accessed over a common PC bus such as PCI-33, PCI-X, FireWire or USB. A suitable bus has to be selected that is fast enough to transfer both data to be processed, and the FPGA reconfiguration bitstream. Since there are such a wide variety of PC bus interface standards, it is also desirable for the PC interface core to be swappable to other standards, depending on what is available on the PC side.

Secondly, a major goal of this work has been to develop a system that allows for fast reconfiguration of the FPGA by the PC. This is with the view that the Proteus Software Platform \cite{bibPSP} running on the PC acts as a supervisor, downloading algorithms in the form of reconfiguration bitstreams to the FPGA as and when desired. This has to be fast because in a processing chain of algorithms, we may have a scenario where data is first processed by a particular algorithm on the FPGA before the results are returned to the PC, and the FPGA has to be reconfigured with the next algorithm in the processing chain to continue processing of that data in the next step. Fast reconfiguration will minimise the delay introduced by this reconfiguration step.

Thirdly, it is desirable to keep the physical hardware design simple, with as low a component count as possible. This will contribute towards one of the goals of keeping the development costs of this platform low. One way in which this can be done is to run both the PC interface core and the algorithm implementation on the same FPGA. However, this will bring about the additional requirement that the part of the FPGA holding the algorithm should be reconfigurable dynamically and independently of the PC interface, even with both on the same chip. To keep the hardware design simple, each hardware board also need not have multiple FPGAs because the software platform allows the concurrent use of FPGAs on multiple boards.
Other reconfigurable platform architectures have been developed, but none satisfies all the requirements given in this section. For example, Blodget et. al \cite{bibXilinxSRP} have designed a self-reconfiguring platform for embedded systems which utilises a soft processor core within the FPGA instead of an external PC. Fong et. al. \cite{bibVirginia} have developed a system that uses the RS-232 port to transfer configuration data to the FPGA, the low transfer speed of which is a limiting factor in reconfiguration performance.

\section{System Architecture}
\label{sectSystemArchitecture}
Considering the requirements set out in Section \ref{sectDesignConsiderations}, the hardware architecture was designed to utilise the PCI bus and the self reconfiguration capability of the Xilinx Virtex II FPGA. 

The PCI bus is commonly found in typical PCs, and allows the hardware to be fully powered from the bus itself. This removes the need for additional power-supply circuits, and avoids electric current separation issues in PCI signalling. In addition, the high-throughput of the bus (132MByte/sec for the 33MHz 32-bit PCI) allows for fast transfer of the reconfiguration bitstream from the PC to the FPGA. The maximum rate at which a bitstream can be transferred over the SelectMap bus to reconfigure a Xilinx Virtex II FPGA is 50MByte/sec, so the PCI bus can easily allow for operation of the SelectMap bus at its maximum speed.

Using the self reconfigurability feature of the FPGA allows one part of the FPGA to hold the PCI interface, while the other holds the downloaded algorithm. This allows the hardware board to remain very simple, removing the need for a separate PCI interface chip. This approach also reduces the number of possible points of system failure caused by physical connections. At the same time, the Xilinx modular design flow \cite{bibModFlow} allows the PCI interface and algorithm parts to be developed and tested independently, even though both are on the same FPGA.

The combination of these features satisfies all the three requirements of using a common PC bus standard, fast FPGA reconfiguration timings, and a simple physical hardware board. The details of the system architecture are explained below.

\subsection{Physical Hardware Design and Initial Bootup Configuration}
\begin{figure}[htb]
\begin{center}
\includegraphics[width=0.4\textwidth]{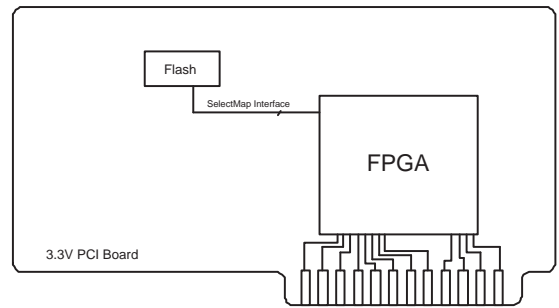}
\caption{Hardware Design}
\label{figHardware}
\end{center}
\end{figure}

As described in the third design consideration of Section \ref{sectDesignConsiderations}, it is desirable for the physical hardware to be kept as simple as possible. The hardware board was therefore designed with only two main components - the FPGA and a Flash RAM chip, as shown in Figure \ref{figHardware}.

The FPGA is placed as close as possible to the physical PCI interface, so that it will satisfy the PCI specification requirements for signal line lengths.

The Flash memory chip holds the FPGA's initial configuration bitstream, which is downloaded to the FPGA through the SelectMap interface upon power-up. This is necessary because SRAM-based FPGAs are volatile and lose their configuration when power is removed. This initial configuration bitstream holds the logic for the PCI interface core and the self-reconfiguration controller.

Once the initial configuration is done, the host PC may download partial bitstreams over the PCI bus to reconfigure the FPGA according to the desired algorithm.

In designing the physical hardware, it was decided that only a single FPGA would be included on each board. Other hardware designs exist that use arrays of FPGAs, but these introduce additional design issues in interchip connectivity, communication protocol specification, and task distribution. With the increase in the density and available logic area of FPGAs in recent years, complex algorithms can now be easily contained in a single FPGA.

Having a single chip solution benefits the algorithm designer in that he will not need to be concerned with the complexity of breaking his design down into various parts for each FPGA, and constraining communications between these parts to the interchip signal lines that have been routed. In an event where more than one FPGA is needed, the Proteus Software Platform can utilize multiple boards concurrently.

\subsection{FPGA Partitioning for Single-Chip Architecture}
\begin{figure}[htb]
\begin{center}
\includegraphics[width=0.4\textwidth]{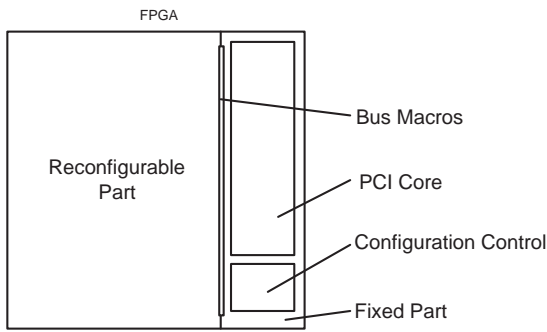}
\caption{Partitioning of the FPGA}
\label{figPartition}
\end{center}
\end{figure}

The use of a single FPGA to hold both the PCI core and the desired algorithm involves partitioning the chip into two modules - a 'fixed' part and a 'reconfigurable' part. The 'fixed' part holds the PCI core and configuration controller, and is defined by the initial full bitstream downloaded from the Flash upon bootup. The 'reconfigurable' part is dynamically configured according to the desired algorithm, via a corresponding partial bitstream download from the PC. This effectively uses the FPGA as a partial self-reconfigurable system, with the fixed part configuration controller internally performing this dynamic reconfiguration operation on the 'reconfigurable' part. Both these parts are linked via Bus-Macros \cite{bibXAPP290}, which provide a well-defined physical interface of signal lines. Figure \ref{figPartition} shows the partitioning of the FPGA for these four main components.

The Xilinx Virtex 2 XC2V3000 was used for the prototype developed. The architecture of the Virtex 2 family requires module areas to span the full height of the chip. Reconfiguration granularity is thus restricted to full columns, and modules correspondingly have only a 1-dimensional flexibility of area placement.

The 'fixed' part is constrained to occupy the right-most columns because the Internal Configuration Access Port (ICAP) in a Virtex 2 FPGA is located at the lower-right corner. The ICAP is used by the Configuration Control logic to reconfigure the 'reconfigurable' part of the chip. The IO pin pads used by the PCI core are accordingly lined up along the right side border of the FPGA.

\subsection{Modular Design Flow}
In implementing algorithms for the reconfigurable part, it will be undesirable and inconvenient if the entire design (including the fixed part) has to be compiled and tested as a whole each time. The Xilinx Modular Design Flow allows each module (the 'fixed' part and the 'reconfigurable' part) to be developed, tested, and debugged independently. Only the reserved module area and the Bus-Macro interface have to remain consistent, and are defined in the common 'top-level' information used in individual module designs. The module area constraints are used in the place-and-route (par) step to define the module boundaries, outside of which logic placement and signal routing is disallowed.

From the individual module design, a 'partial bitstream' can be created that represents the configuration information for a single module. This 'partial bitstream' created for the 'reconfigurable' part is dynamically downloaded from the PC to reconfigure the corresponding portion on the FPGA, according to the desired algorithm.

The 'full bitstream' derived from the design of the 'top-level' and 'fixed' part is stored on the Flash for the initial boot-up configuration of the FPGA.

Although Xilinx has stated \cite{bibXAPP290} that it is not possible to use the PCI core in the Modular Design Flow, this work has shown that it is actually possible to do so.

\subsection{Self-Reconfiguration using the Internal Configuration Access Port (ICAP)}
The Configuration Controller of the 'fixed' part uses the Internal Configuration Access Port (ICAP) of the Xilinx Virtex 2 FPGA to perform self-reconfiguration. This process is carried out without influence of the host PC's CPU.

The Configuration Controller obtains the reconfiguration data from a memory location shared with the PC. This avoids having to place a restriction on the configuration bitstream size based on the available on-chip FPGA RAM. In other self-reconfigurable systems \cite{bibVirginia} \cite{bibXilinxSRP}, a limited amount of on-chip memory is used to store the reconfiguration data. This results in the self-reconfiguration process taking a long time because of the need to iteratively load portions of the configuration bitstream and incrementally reconfigure the FPGA. 

A second limitation is in the throughput of the link over which the reconfiguration bitstream is sent - this often becomes a bottleneck in the speed of reconfiguration if a low throughput connection is used, such as in \cite{bibVirginia} where an RS232 serial line is used.

In this work, the bus-master capability of the PCI core allows for direct use of the PC's memory, which is large enough to hold an entire configuration bitstream. The data is transferred in a continuous stream over the PCI bus to the ICAP, removing the need for incremental reconfiguration and offering the fastest possible speed of reconfiguration over the high-throughput PCI bus.

The ICAP uses the Select Map Bus protocol, so the Configuration Controller has to act as a bridge between that and the PCI bus. To provide for clock independence between these parts, on-chip dual-port Block RAMs (BRAMs) are used as a buffer. This technique avoids the need to exchange ownership of a shared RAM space between the PCI Interface and the Select Map Interface as done in our previous work \cite{bibRaw}, thereby minimizing the latency time in accessing the ICAP. The prototype implementation uses two BRAMs (on a Virtex 2 XC2V3000) for a total buffer size of 256x32bit.

Internally, the Configuration Controller works with 32-bit items to match the width of the PCI bus. A multiplexer segments each of these 32-bit items into four single byte items, for passing on to the Select Map bus that is 8-bit wide. In the event where the PCI bus fails to deliver sufficient data to the buffer in time, the controller 'pauses' the operation by stopping the configuration clock. It is resumed once data is available again.

\subsection{Fixed Part Design}
The PCI Interface is used for two purposes: to transfer partial reconfiguration bitstreams to the FPGA, and to exchange data with the algorithm downloaded to the FPGA. Both of these transfers can use the bus-master feature of the PCI bus, which allows direct shared access to the PC's RAM to obtain data. This ensures that the design on the FPGA is not limited by the small amount of on-chip RAM, and provides for the fastest possible direct data transfer from the PC.

\begin{figure}[htb]
\begin{center}
\includegraphics[width=0.5\textwidth]{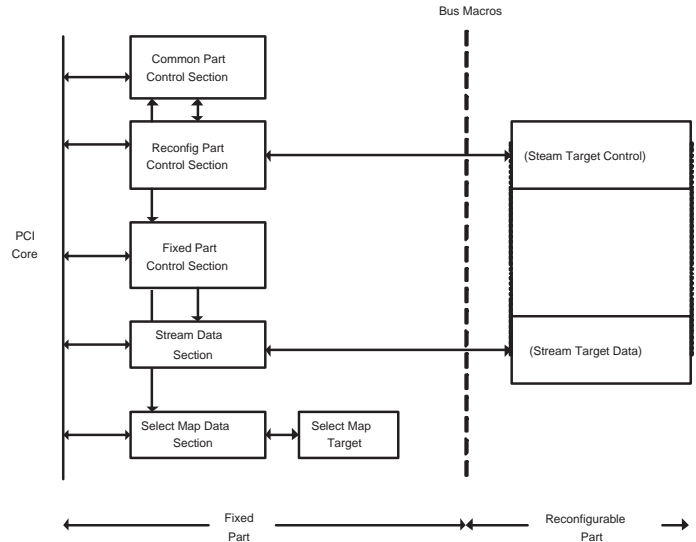}
\caption{Fixed Part Design}
\label{figFixedPartDesignConcept}
\end{center}
\end{figure}

The design of the fixed part which allows for this is shown in Figure \ref{figFixedPartDesignConcept}.

The major components of the fixed part include the Fixed Part Control Section, Reconfig Part Control Section, Common Part Control Section, Stream Data Section, and Select Map Data Section.

The fixed part contains the control logic necessary for busmaster mode access of the PCI bus. It also handles all communication with the device driver, and transports data to/from the reconfigurable part. The fixed part acts as a bridge between the packet oriented transfer of the shared PCI bus and the continuous data stream transfer with the reconfigurable part.

The fixed part is split into three control sections - the Common Part Control, Reconfig Part Control, and Fixed Part Control, and two data sections - Stream Data and Select Map Data. The Common Part Control holds PCI access specific functions and contains the interrupt generation logic. The Reconfig Part Control allows the reconfigurable part to easily implement static registers that can be individually addressed, read from, and written to. Using these registers, the reconfigurable part may obtain setup values from the PC side and report back status information. The reconfigurable part may also invoke interrupts via the fixed part design. The Fixed Part Control section controls the PCI bus / reconfigurable part data transfer tasks. The Data sections each hold Dual-Port Block RAMs (BRAMs) to buffer data between the PCI bus and the reconfigurable part. The Dual-Port feature allows the BRAMs to be accessed from different clock domains on each side. One port is accessed from the PCI side and driven with the PCI clock, while the other port is accessed from the reconfigurable part and driven with the clock from that. A buffer size of 256x32bits for each stream port has been found sufficient to contain the access latency of the PCI bus.

\begin{figure}[htb]
\begin{center}
\includegraphics[width=0.5\textwidth]{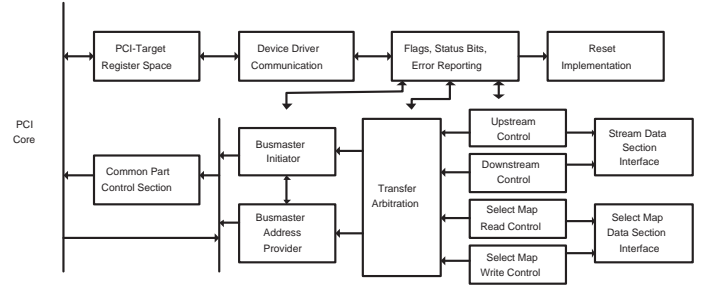}
\caption{Fixed Part Control Section}
\label{figFixedPartControl}
\end{center}
\end{figure}

The functional blocks of the Fixed Part Control Section is shown in Figure \ref{figFixedPartControl}.

The Fixed Part Control contains a Busmaster Initiator and a Busmaster Address Provider. The Busmaster Initiator starts bus-master data transfers with the allocated RAM on the PC, over the PCI bus. This is done at each transfer event, which are triggered on the fill status of the internal BRAM buffers. The Busmaster Address Provider keeps track of the PCI addressing, and is needed because a transfer on the shared PCI bus can be interrupted at any time. When such an interruption happens the transfer has to be restarted at the next address, after the bus is available again.

The Transfer Arbitration block schedules the utilization of the PCI Interface by the four possible stream targets - Upstream, Downstream, Select Map Read, and Select Map Write. All the four targets utilize the same PCI Interface, so the Transfer Arbitration block uses a simple scheduling algorithm to equally allocate data transfer requests - one target is not allowed to request the PCI Interface twice in a row if multiple requests are pending. Each stream target has a corresponding Control section, which generate data transfer requests and monitor the status of the BRAM buffers in the Data sections.

\section{Experimental Results}
\label{sectExperiments}

\begin{figure}[htb]
\begin{center}
\includegraphics[width=0.4\textwidth]{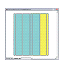}
\caption{Regions of the FPGA allocated for the fixed part (yellow) and the reconfigurable part (blue)}
\label{figFloorPlanner}
\end{center}
\end{figure}

\begin{figure}[htb]
\begin{center}
\includegraphics[width=0.4\textwidth]{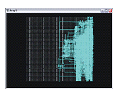}
\caption{Resources used by the Fixed Part}
\label{figScreenCapture}
\end{center}
\end{figure}

The fixed part design as described in the previous section was compiled for a Xilinx XC2v3000ff1152-4 FPGA using the Xilinx ISE 6.2. Figure \ref{figFloorPlanner} shows a screen capture of the ISE FloorPlanner with the regions of the FPGA allocated for the fixed part in yellow, and that for the reconfigurable part in blue. Figure \ref{figScreenCapture} is a screen capture from the ISE FPGA Editor, showing the resources used by the fixed part. A total of 2211 slices (out of a total of 14336) and 3 BRAMs are used, taking up about 15\% of the XC2V3000.

\section{Conclusion}
\label{sectConclusion}

This paper has presented a reconfigurable computing platform hardware architecture that satisfies the design requirements of being PC based, allowing for fast reconfiguration over the PCI bus, and simplicity of physical hardware design.

\section*{Acknowledgments}
We gratefully acknowledge the funding support provided by the Ngee Ann Kongsi (Singapore) and Ngee Ann Polytechnic's Innovation \& Enterprise Office. Special thanks to the rest of the Proteus Team - Philip Wong, Kelvin Lim and Kelly Choo.

\nocite{*}
\bibliographystyle{IEEE}

%

%

\end{document}